\documentclass[twocolumn]{aastex631}

\usepackage{color}
\usepackage{amsmath}
\usepackage{xcolor}
\usepackage{mathrsfs}
\usepackage{natbib}
\usepackage{gensymb}

\shortauthors{Bryan et al.}
\shorttitle{HD106906\lowercase{b} Obliquity}

\begin{document}

\title{Obliquity Constraints on the Planetary-mass Companion HD 106906 \lowercase{b}}

\author{Marta L. Bryan}
\affiliation{Department of Astronomy \\
501 Campbell Hall \\
University of California Berkeley\\
Berkeley, CA 94720-3411, USA}

\author{Eugene Chiang}
\affiliation{Department of Astronomy \\
501 Campbell Hall \\
University of California Berkeley\\
Berkeley, CA 94720-3411, USA}

\author{Caroline V. Morley}
\affiliation{Department of Astronomy\\
The University of Texas at Austin\\
Austin, TX 78712, USA}

\author{Gregory N. Mace}
\affiliation{Department of Astronomy\\
The University of Texas at Austin\\
Austin, TX 78712, USA}

\author{Brendan P. Bowler}
\affiliation{Department of Astronomy\\
The University of Texas at Austin\\
Austin, TX 78712, USA}

\begin{abstract}

We constrain the angular momentum architecture of HD 106906, a 13 $\pm$ 2 Myr old system in the ScoCen complex composed of a compact central binary, a widely separated planetary-mass tertiary HD 106906 b, and a debris disk nested between the binary and tertiary orbital planes. We measure the orientations of three vectors: the companion spin axis, companion orbit normal, and disk normal.  Using near-IR high-resolution spectra from Gemini/IGRINS, we obtain a projected rotational velocity of $v\sin i_p$ = 9.5 $\pm$ 0.2 km/s for HD 106906 b. This measurement together with a published photometric rotation period implies the companion is viewed nearly pole-on, with a line-of-sight spin axis inclination of $i_p$ = 14 $\pm$ 4$\degree$ or 166 $\pm$ 4$\degree$. By contrast, the debris disk is known to be viewed nearly edge-on. The likely misalignment of all three vectors suggests HD 106906 b formed by gravitational instability in a turbulent environment, either in a disk or cloud setting.

\end{abstract}


\section{Introduction}

Planetary obliquity measurements inform our understanding of how planets form and evolve.  A planetary obliquity is the mutual inclination between the planet's spin axis and its orbit normal.  Up until last year, only solar system planets had measured obliquities. In this single system we find a wide range of orientations --- Uranus is on its side, Venus is upside-down, and Earth is tilted by 23$^\circ$ which gives us our seasons. From these spin rates and directions we can infer planet formation histories. The terrestrial and ice giant planets likely experienced giant impacts, tidal friction, and gravitational forcing  \citep[e.g.][]{Dobrovolskis1980,Lissauer1991,Laskar1993,Touma1993,Correia2006,Schlichting2007,Reinhardt2019}. While the spin axes of Jupiter and Saturn may have initially both been aligned with the angular momentum of the broader circumstellar disk, secular spin-orbit resonances driven by orbital migration have been invoked to explain Saturn's 27$^\circ$ obliquity \citep[e.g.][]{Ward2004,Nesvorny2018}.

These processes and others can apply to exoplanets. Theoretical work shows that obliquities can be excited through secular spin-orbit resonances created by planet-planet or planet-disk interactions \citep[e.g.][]{Millholland2019,Millholland20192,Li2021}. Kozai-Lidov oscillations from an external perturber can be expected to produce significant planetary and stellar obliquities \citep[e.g.][]{Storch2014,Martin2014}. Spin axes may also be tilted at the time of formation: turbulence in self-gravitating disks is expected to produce a dispersion of spin axis directions for fragmenting clumps \citep{Bryan2020,Jennings2021}.  

A planet's obliquity can be constrained from three observables: the projected rotation rate $v\sin i$ of the planet, its photometric rotation period $P_{\rm rot}$, and a 3D orbit.  Combining $v\sin i$, $P_{\rm rot}$, and a radius estimate yields the line-of-sight spin axis inclination of the planet, and the orbit plane gives the orbital inclination.  At present, the only planets amenable to these measurements are $\sim$25 young super-Jupiters discovered by direct imaging campaigns \citep[e.g.][]{Bowler2016}.  Because these objects are young ($\lesssim$ 100 Myr old) and massive ($\sim$10 -- 20 M$_{\rm Jup}$), they are relatively bright, and their large separations from their host stars ($\gtrsim$ 50 AU, $\gtrsim$ 1 arcsec) help ensure that their fluxes are not buried beneath the glare of their host stars.  It is thus feasible to extract spectra and light curves for the planets themselves, thereby measuring $v\sin i$ and $P_{\rm rot}$.

However, it is exceptionally rare to obtain all three of these observables for a single object.  To date $\sim$15 planetary-mass companions have measured rotation rates (see Table 3 in \citet{Bryan2020_2}). Only companions 2M0122 b and VHS 1256-1257 b have both a measured $v\sin i$ and $P_{\rm rot}$. Some objects with measured $P_{\rm rot}$ are too faint to extract a spectrum and measure $v\sin i$. Others that have measured $v\sin i$'s do not have detectable rotational modulations in their light curves, precluding a $P_{\rm rot}$ constraint. Some of these companions, such as VHS 1256-1257 b, are so far from their host stars that constraining the 3D orbit is not feasible. Prior to this work there was only one system with all three pieces in hand.

\citet{Bryan2020} placed the first constraint on the obliquity of a planetary-mass object outside the solar system.  This study focused on the 120 Myr old system 2MASS J01225093--2439505, which comprises a 0.4 M$_{\odot}$ host star with a 12--27 M$_{\rm Jup}$ companion (hereafter 2M0122 b) orbiting at 52 AU \citep{Bowler2013}. Line-of-sight inclinations for the planetary spin, stellar spin, and orbital angular momentum vectors were measured using projected rotational velocities $v\sin i$'s for the star and companion, rotation periods $P_{\rm rot}$'s for the star and companion, and an astrometric orbit for the companion. There is evidence that the true stellar obliquity is small and the true companion obliquity is large, although there are large uncertainties because of the unknown orientation of the spin axes in the sky plane.  A promising scenario that could account for these mutual inclinations is formation via instability in a gravito-turbulent disk, wherein turbulent eddies of gas spinning in a variety of directions collapse under their own self-gravity, yielding a range of obliquities for the resulting objects  \citep{Bryan2020,Jennings2021}.

Here we present constraints on a second extrasolar planetary-mass companion obliquity.  We study HD 106906, a 13 $\pm$ 2 Myr old system with a central close binary (masses 1.37 and 1.34 M$_{\odot}$, orbital period 49.233$\pm$0.001 days, and eccentricity 0.669$\pm$0.002), orbited by an 11.9$^{+1.7}_{-0.8}$ M$_{\rm Jup}$ companion at a projected separation of $\sim$737 AU \citep{Bailey2014,Nguyen2021}.  This system also hosts an asymmetric debris disk with a vertically thin eastern side that extends to over 550 AU, and a vertically thick western side that reaches a radius of $\sim$ 370 AU \citep{Kalas2015,Lagrange2016}.  We seek to constrain three angular momentum vectors: the planetary spin axis, planetary orbit normal, and debris disk normal.  We do not include the binary star system in our analysis as the angular momentum vectors for the binary orbit and stellar spins are unknown.  While recent work has shown that close, circular binaries have orbital planes that are more likely to be aligned with the planes of their circumbinary debris disks \citep{Czekala2019}, the central binary in HD 106906 has an orbital period that is too long and an eccentricity that is too high to safely make the assumption that the binary plane and the debris disk plane are  coplanar.

This paper is organized as follows.  In Section 2 we describe our high-resolution spectroscopic observations with IGRINS/Gemini.  Section 3 lays out measurements of the line-of-sight inclinations of the planetary spin, orbital, and disk angular momentum vectors, and gives constraints on the true 3D angles between each pair. We discuss what physical scenarios could account for these constraints in Section 4, and present our conclusions in Section 5.

\section{Observations}

\subsection{IGRINS/Gemini High-Resolution Spectroscopy}
Observations of HD 106906 b with the Immersion Grating Infrared Spectrometer \citep[IGRINS:][]{yuk2010,park2014} on the Gemini South telescope \citep{mace2018spie} were completed 2020 February 04, 07, 08, 09 UT as part of program GS-2020A-Q-135 (PI: Bryan). These observations simultaneously covered \textit{H} and \textit{K} bands from $\sim$1.45 -- 2.52 $\mu$m.  We observed the \textit{K} = 15.5 mag companion with individual image exposure times of 1528 seconds. All observations were taken with a slit orientation perpendicular to the 307.3$\degree$ position angle between the host star and the companion (angular separation 7.1$"$) in order to prevent a flux gradient across the slit.  On 2020 February 04 UT we acquired three pairs of AB nodded exposures, amounting to three epochs of observation in $\sim$2.5 hours of on-source integration time. On the nights of 2020 February 07, 08, and 09 UT ABBA-nodded exposures were combined, producing three additional epochs. In total, there were six epochs of observation from four nights in early February 2020.

\section{Analysis}

\subsection{Measuring $v \sin i$ for HD 106906 b}

We reduce all data with the IGRINS Pipeline Package \citep[PLP;][]{lee2016}. The package uses AB pairs of slit-nodded spectra to sky subtract and then optimally extract the target flux. Wavelength calibration is carried out using both OH sky emission and telluric absorption from the A0V star observed immediately before or after the target. The A0V star also serves as a telluric standard, and when the target spectrum is divided by the A0V spectrum the target is corrected for both telluric absorption and the instrument profile. The final product of the PLP is a wavelength-calibrated spectrum of the target star, with flux in counts, and corresponding signal-to-noise spectrum. While reduced spectra were produced across the wavelength range $\sim$1.45--2.52 $\mu$m, we only consider the \textit{K}-band spectra in subsequent analyses given the low signal-to-noise ratio (SNR) of the \textit{H}-band spectrum. This companion is brightest in \textit{K}-band (1.85 -- 2.52 $\mu$m), although we find that some \textit{K}-band orders are also unusable due to low SNR.

Instrumental resolution and rotation both produce line broadening, and these two sources of broadening are degenerate. It is thus important to accurately measure the resolution in order to accurately measure $v\sin i$. Using observed standard star spectra, we first select four IGRINS orders spanning wavelengths 2.293--2.325~$\mu$m, 2.236--2.267~$\mu$m, 2.182--2.212~$\mu$m, and 2.105--2.135~$\mu$m.  For each of the six epochs of data, we use the \texttt{molecfit} routine, which simultaneously fits a telluric model and an instrumental profile defined by a single Gaussian kernel, to the spectrum \citep{smette_molecfit2015, kausch_molecfit2015}.  We take the median of these 24 instrumental resolution measurements as the instrumental resolution to use in our analysis, and the standard deviation to be the uncertainty. From this analysis we find \textit{R} = 48599$\pm$1514.  We also check whether the resolution changes significantly within an order.  We select one of the epochs taken on UT February 04 2020, and divide each of the four orders into five parts (each $\sim$ 400 pixels across).  We find that the resulting instrumental resolution values within each order are consistent with each other and with the global resolution measurement.

In the wavelength-calibrated and telluric-corrected output spectra from the IGRINS reduction routine, we remove artifacts from strong sky lines that manifested as spikes in the data.  In addition, we find that the short wavelength end of each spectrum contains less flux as the instrument blaze falls off.  We cut the leftmost 20 -- 70 pixels off of each reduced spectral order, where the cutoff value grew with increasing order number (decreasing wavelength).  

With these spectra of HD 106906 b, we want to measure the amount of line broadening due to the rotational velocity. To do so, we calculate the ``data'' cross-correlation function (CCF) between the observed spectrum and a model atmosphere, where the model has been broadened to the instrumental resolution.  We use an atmospheric model from the Sonora model grid \citep[Morley et al. in prep.]{Marley2021}. These models are calculated assuming that the atmosphere is in radiative--convective and chemical equilibrium, following the approach of \citet{Marley1999, Saumon2008, Morley2012}, with updated chemistry and opacities as described in \citep{Marley2021}. We assume $T_{\rm eff}$ = 1820 K and $\log(g)$ = 4.0 for HD 106906 b, following measurements of $T_{\rm eff}$ = 1820 $\pm$ 240 K and $\log(L_{\rm bol}/L_{\odot})$ = -3.65 $\pm$ 0.08 using medium resolution spectra from VLT/SINFONI \citep{Daemgen2017} and converting $\log(L_{\rm bol}/L_{\odot})$ and system age to $\log(g)$ using hot-start evolutionary models \citep{Burrows1997}. The Sonora models generated for HD 106906 b have solar metallicity and solar carbon-to-oxygen ratio (C/O), and include silicate, iron, and corundum clouds with a sedimentation efficiency $f_{\rm sed}=2$ as described in \citet{Ackerman2001}.

We compare this ``data'' CCF to ``model'' CCFs, where each model is calculated by cross-correlating a model atmosphere broadened by the instrumental line profile with that same model additionally broadened by some rotation rate and offset by a radial velocity (RV).  We perform this comparison in a Bayesian framework using MCMC to fit for three free parameters:  $v\sin i$, RV, and instrumental resolution.  We use uniform priors on $v\sin i$ and RV.  For the instrumental resolution, we use a Gaussian prior with a mean of 48599 and standard deviation of 1514 to account for uncertainties on the measured resolution.  

The log-likelihood function used in our MCMC framework is given by
\begin{equation}
\log{L} = \sum_{i=1}^{n} -0.5\bigg(\frac{m_i - d_i}{\sigma_{i}}\bigg)^2,
\label{eq:loglike}
\end{equation}
\noindent where $d$ is the ``data'' CCF, $m$ is the ``model'' CCF, and $\sigma_i$ is the CCF error at position $i$.  We calculate uncertainties on the ``data'' CCF using the jackknife resampling technique.  In this case, uncertainties are given by 
\begin{equation}
\sigma_{\rm{jackknife}}^2 = \frac{(n-1)}{n} \sum_{j=1}^{n} {(x_j - x)}^2,
\label{eq:jackknife}
\end{equation}
\noindent where $n$ is the total number of samples.  We define a sample as one epoch of data -- there are six for HD 106906 b.  $x_j$ is the ``data'' CCF calculated using all epochs of data except the $j$th epoch, and $x$ is the ``data'' CCF calculated using all epochs of data.

\begin{figure}[h]
\centering
\includegraphics[width=0.5\textwidth]{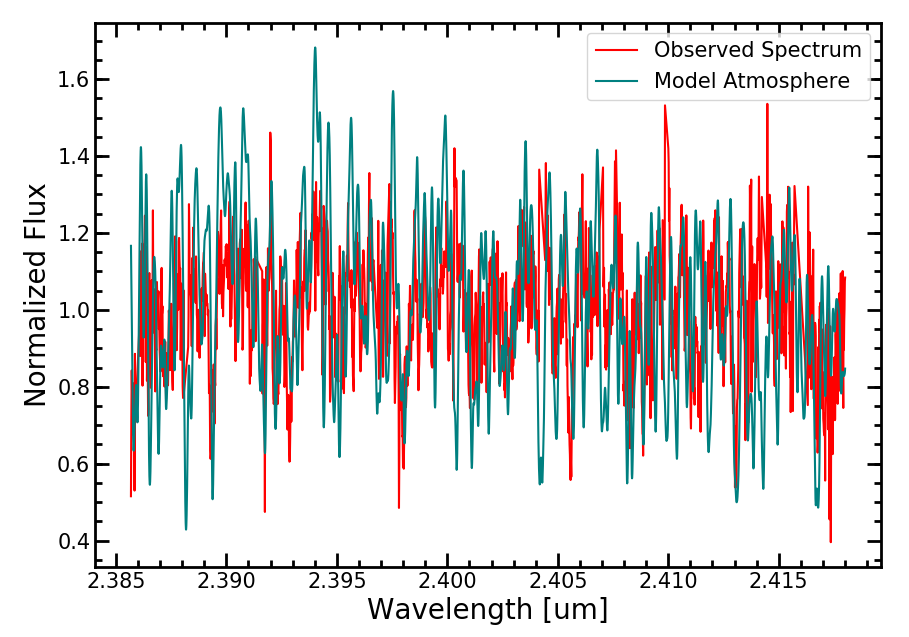}
\includegraphics[width=0.5\textwidth]{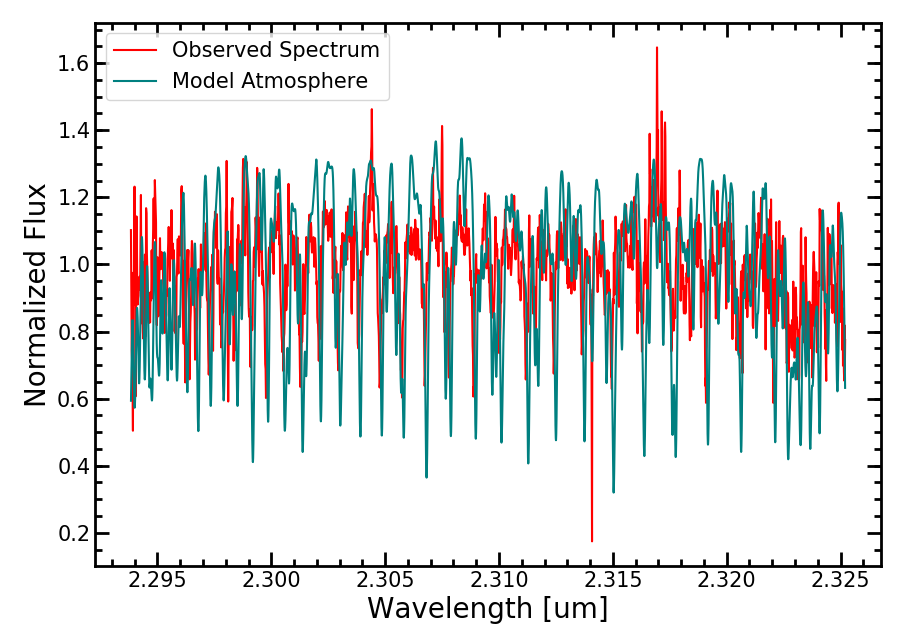}
\caption{Orders 4 (top) and 7 (bottom) spectra for HD 106906 b (red), overplotted with a model atmosphere broadened to the best-fit rotation rate.}
\label{fig: order spectra}
\end{figure}

\begin{figure}[h]
\centering
\includegraphics[width=0.5\textwidth]{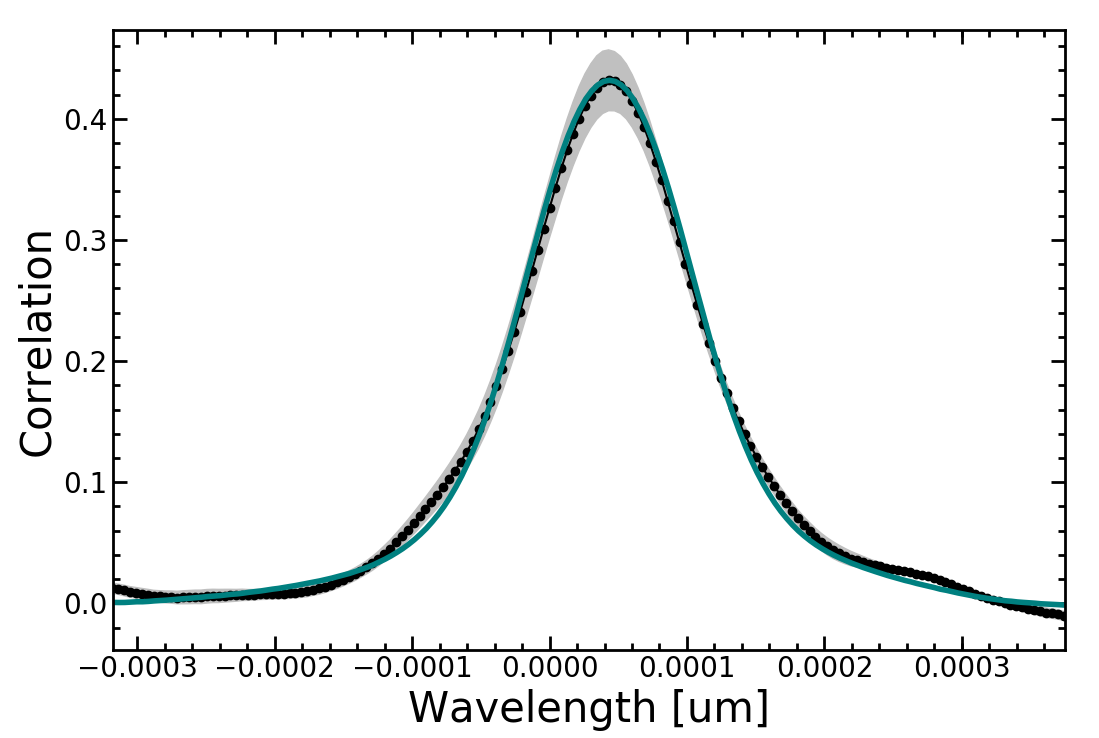}
\caption{Cross correlation function between orders 4--19 (1.99 -- 2.42 $\mu$m) of the observed spectrum with a model atmosphere broadened to the instrumental resolution (black points), shown with 1$\sigma$ uncertainties shaded in gray calculated using jackknife resampling technique.  The cross-correlation function between a model atmosphere broadened to the instrumental resolution, and that same model additionally broadened by the best fit rotation rate and shifted by the best fit velocity offset is shown in teal.}
\label{fig: best fit CCF}
\end{figure}

\begin{figure}[h]
\centering
\includegraphics[width=0.5\textwidth]{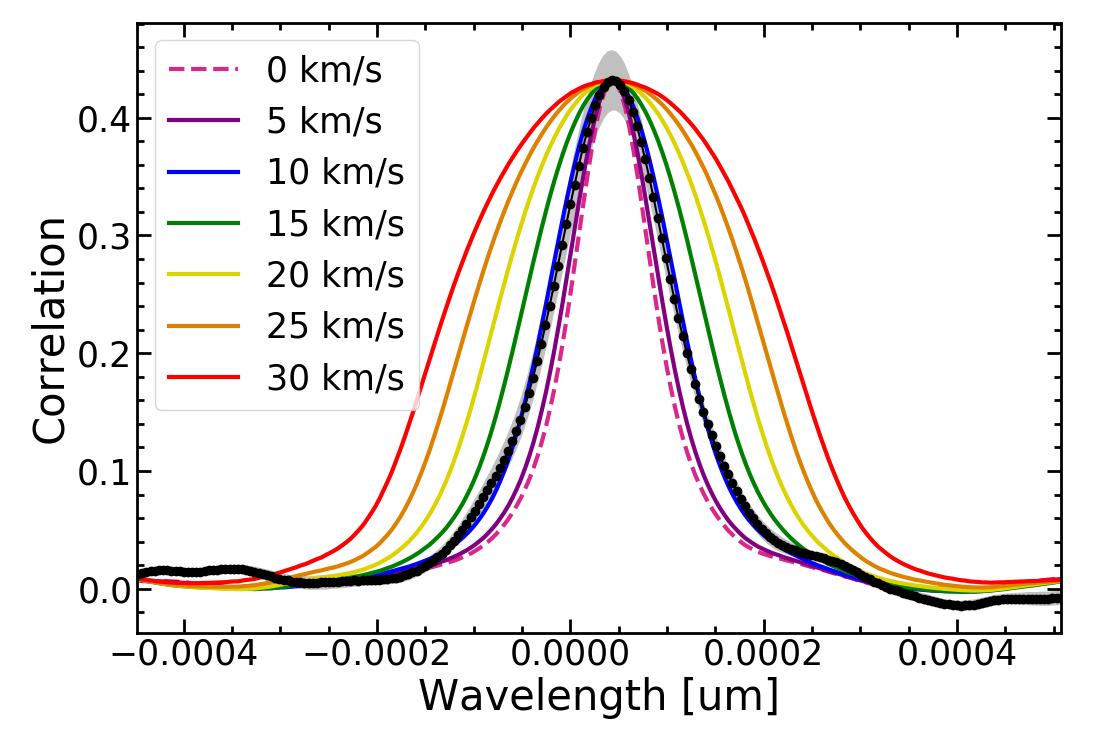}
\caption{Cross correlation function between orders 4--19 (1.99 -- 2.42 $\mu$m) of the observed spectrum with a model atmosphere broadened to the instrumental resolution (black points), shown with 1$\sigma$ uncertainties shaded in gray calculated using jackknife resampling technique. The cross-correlation functions between a model atmosphere broadened to the instrumental resolution, and that same model additionally broadened by a series of rotation rates (0, 5, 10, 15, 20, 25, 30 km/s) are shown in color.}
\label{fig: rainbow CCF}
\end{figure}

Before undertaking a joint fit of multiple orders to determine $v\sin i$, we first consider each order in \textit{K}-band individually.  We compute ``data'' CCFs for 22 orders spanning wavelengths 1.85 -- 2.42 $\mu$m (the first three spectral orders extended to longer wavelengths than those covered by our models), and determine the significance of the peak (if present) by calculating the ratio of the peak height to the standard deviation of the CCF outside the central peak. We exclude orders with peaks with less than 5$\sigma$ significance.  This cut leaves us with 15 orders running from 1.99 -- 2.42~$\mu$m (excluded orders have significant telluric features and lower SNR spectra).  We fit each order individually to get independent estimates for $v\sin i$, and find that they are consistent within their uncertainties.  We then perform a joint fit, calculating a ``data'' CCF using all 15 orders and fitting models across that entire 1.99 -- 2.42 $\mu$m wavelength range.  The measured projected rotation rate for HD 106906 b is $v\sin i$ = 9.5 $\pm$ 0.2 km/s (see Figures \ref{fig: order spectra}-\ref{fig: rainbow CCF} for reference).

We now consider how our modeling assumptions could impact the measured $v\sin i$.  First we test our choice of $T_{\rm eff}$ and $\log(g)$.  From \citet{Daemgen2017} we have $T_{\rm eff}$ = 1820$\pm$240 K, and we converted $\log(L_{\rm bol}/L_{\odot})$ to $\log(g)$ = 4.0$\pm$0.5 using hot start evolutionary models \citep{Burrows1997}.  We take the 1$\sigma$ errors on these values and generate four models:  (1580K, 3.5 dex), (1580K, 4.5 dex), (2060K, 3.5 dex), and (2060K, 4.5 dex). We calculate new $v\sin i$ values with each of these models to test the possible impact of the measured uncertainties on our adopted $T_{\rm eff}$ and $\log(g)$, and found that all $v\sin i$ values were consistent with our original measurement at the $\leq 2\sigma$ level (see Table \ref{tb:model}).  

Another assumption we make when generating atmospheric models is a C/O ratio. In our original model we assume a solar (0.54) C/O value, and here we test three additional ones:  0.25$\times$solar, 0.5$\times$solar, and 1.5$\times$solar. When we implement these models in our MCMC framework, we find that resulting $v\sin i$ values are consistent with the original value to $\leq 0.5\sigma$ for the sub-solar C/O ratios, and differ by 2.3$\sigma$ for the 1.5$\times$solar model (see Table \ref{tb:model}).  While not significant, this tentative offset in $v\sin i$ due to higher C/O suggests that abundance assumptions can become important for $v\sin i$'s given our measurement precision of 0.2 km/s.

Finally, we explore the impact that uncertainties in pressure broadening can have on the measured rotational velocity.  Pressure broadening is a degenerate effect along with instrumental broadening and rotational line broadening -- higher pressure broadening with the same instrumental resolution leads to less rotational line broadening and a correspondingly smaller $v\sin i$. To test our pressure broadening assumptions, we run two models with modified molecular opacities, where molecular cross sections were 10$\times$ and 0.1$\times$ the actual pressure for the whole profile.  This simulates a scenario where pressure broadening parameters that are used to create the molecular cross sections are off by an order of magnitude. Collision-induced opacity of hydrogen and helium is treated separately for all models, using the standard pressure for each layer.  When we recalculate $v\sin i$ values using these new models, we find that for the 0.1$\times$ model (which under predicts the amount of pressure broadening) the resulting rotation rate is only 1.0$\sigma$ away from the original value, and the 10$\times$ model (which over predicts pressure broadening) produces a $v\sin i$ that is 1.7$\sigma$ lower (see Table \ref{tb:model}). Both values are consistent with the original rotation rate measurement. 

\begin{deluxetable}{cc}
\tablecaption{Model Tests and Resulting $v\sin i$'s \label{tb:model}}
\tabletypesize{\footnotesize}
\tablehead{
  \colhead{Model} & 
  \colhead{$v\sin i$} 
}
\startdata
Original & 9.53$\pm$0.24 km/s  \\
1580 K, 3.5 dex & 9.88 (+0.25 -0.24)\\
1580 K, 4.5 dex & 8.85 $\pm$ 0.22\\
2060 K, 3.5 dex & 9.89 (+0.23 -0.26)\\
2060 K, 4.5 dex & 9.20 (+0.22 -0.21)\\
0.25$\times$ solar C/O & 9.63 (+0.24 -0.23)\\
0.5$\times$ solar C/O & 9.69 (+0.28 -0.24)\\
1.5$\times$ solar C/O & 8.73 (+0.25 -0.20)\\
0.1$\times$P & 9.87 (+0.20 -0.22)\\
10$\times$P & 8.84 (+0.33 -0.28)
\enddata
\end{deluxetable}

\subsection{Measuring P$_{rot,p}$ for HD 106906 b}

Periodic features in substellar light curves can be produced by cloud patchiness or planetary-scale waves, which manifest as longitudinal bands produced by zonal circulation \citep{Apai2017,Apai2021}. Lower gravity objects typically have higher variability amplitudes and higher intrinsic variability rates \citep{Metchev2015,Vos2019}. This observed variability appears to be impacted by viewing geometry -- more highly inclined objects (i.e. closer to pole-on) have more attenuated brightness changes \citep{Vos2017}.

The photometric rotation period for HD 106906 b was published by \citet{Zhou2020}.  The authors used the Hubble Space Telescope Wide Field Camera 3 (WFC3) near-IR channel in time-resolved direct imaging mode to observe HD 106906 b in three bands:  F127M, F139M, and F153M.  Using techniques such as two-roll differential imaging and hybrid point-spread function modeling yielded $\sim$1$\%$ precision in the light curves across all three bands.  Fitting the light curve with a sinusoid results in a period of 4.1$\pm$0.3 hours and an amplitude of 0.49$\pm$0.12$\%$.

\citet{Zhou2020} present several caveats to the rotation period measurement.  First, they find only marginal evidence of variability, with a significance of 2.7$\sigma$.  Because of this tentative detection, when fitting the light curves the authors applied a strict sinusoidal shape to the photometric modulations, and could not investigate whether the light curve could have multiple peaks.  Previously, \citet{Apai2017} found that for 3 L/T transition brown dwarfs with high signal-to-noise data and extremely long baselines ($>1$ year), the power spectra of their light curves produced peaks at both the full rotation period of the object as well as half the rotation period.  A more recent analysis of long baseline high SNR photometry of Luhman 16 A and B found a similar result, with peaks at both the full and half rotation period \citep{Apai2021}. While this raises the question of whether the detected 4 hour rotation period of HD 106906 b is the full or half rotation period, both higher quality and theoretical light curves of brown dwarfs show that the full rotation period dominates the signal in the power spectrum for a given light curve \citep{Zhang2014,Apai2017,Apai2021}.  In addition, periodicity in the light curves of Jupiter and Neptune correspond to their full rotation periods \citep{Karalidi2015,Simon2016,Ge2019}.  We thus assume that the full rotation period of HD 106906 b is 4.1$\pm$0.3 hours.  

Another caveat to this rotation period measurement is that photometric modulations for HD 106906 b are only detected in the bluest band (F127M), and not in the other two bands.  However, for the majority of substellar objects, rotational modulations are wavelength dependent and have higher amplitudes at shorter (bluer) wavelengths \citep[e.g.][]{Apai2013,Yang2015,Zhou2016,Zhou2019}.  Assuming a similar wavelength dependence for the photometric modulations in HD 106906 b as measured for 2M1207 b \citep{Zhou2016}, the closest spectral type young companion analog to HD 106906 b with detected modulation, the authors predict that the modulation amplitude for the redder bands would have been too small for them to observe.  The detection of modulation in only the bluest band is therefore consistent with low overall amplitude variability and wavelength dependent modulations.

\subsection{Measuring i$_{o}$}
\label{sec:measuring i_orbit}

Given the wide projected separation of HD 106906 b (737 AU), detecting orbital motion and placing constraints on orbital parameters requires long baseline precision astrometry \citep[e.g.][]{Bowler2020}.  Recently, \citet{Nguyen2021} detected orbital motion of this companion using 14 years of astrometry measurements between 2004 and 2017. All data were taken with HST using a combination of the Advanced Camera for Surveys (ACS), the Space Telescope Imaging Spectrograph (STIS), and the Wide Field Camera 3 (WFC3).  By cross-registering background star locations in the HST images with the Gaia astrometric catalog, \citet{Nguyen2021} calculated astrometry to high precision (at the sub-pixel level).  To obtain constraints on the orbital parameters of HD 106906 b, the authors used the open-source Python package \texttt{orbitize!}\footnote{https://github.com/sblunt/orbitize} \citep{Blunt2019} to perform an orbit fit to the assembled astrometric measurements.  The line-of-sight orbital inclination $i_o$ from these fits is 56$^{+12}_{-21}$$\degree$. In this paper we use the posterior distribution shown in Fig. 10 of \citet{Nguyen2021} when incorporating $i_o$ in our analyses.

\subsection{Measuring i$_{d}$}
\label{sec:measuring i_orbit}

Scattered light imaging of the debris disk in the HD 106906 system with both GPI and SPHERE constrained the morphology of the disk \citep{Kalas2015,Lagrange2016}.  In \citet{Lagrange2016}, the disk detection with SPHERE was modeled using the GRATER code \citep{Augereau1999} as an optically thin, inclined ring centered on the host binary.  The authors assumed a dust density distribution that peaks at radius $r_0$ and has a power law slope of $\alpha_{in}$ inside of $r_0$ and $\alpha_{out}$ outside of $r_0$.  In addition to $r_0$ and $\alpha_{out}$, fitted model parameters include the inclination of the disk $i_d$, the position angle (PA), a scaling factor to match the total flux of the disk, and the Henyey-Greenstein coefficient $g$, which quantifies how anisotropic the scattering is.  With this modeling, the authors find a disk inclination $i_d$ = 85.3$\pm$0.1$\degree$.  In \citet{Kalas2015}, the authors estimate a disk inclination of $i_d$ $\sim$ 85$\degree$ by assuming the disk is circular and translating the disk aspect ratio from their fitted semi-major and minor axes to line-of-sight inclination. Because it is unclear whether the disk is rotating in a prograde or retrograde fashion, there is a degeneracy in $i_d$ and $\Omega_d$ pairs, where $\Omega_d$ is the position angle (PA) of the ascending node.  Thus angles ($i_d$ = 85$\degree$, $\Omega_d$ = 104$\degree$) and ($i_d$ = 95$\degree$, $\Omega_d$ = 284$\degree$) are equally likely.  In this paper, we assume a bimodal distribution for $i_d$, with $i_d$ = 85.3$\pm$0.1$\degree$ and $i_d$ = 94.7$\pm$0.1$\degree$ defining the two Gaussian distributions.

\subsection{Measuring $i_{p}$}

We combine $P_{\rm rot}$ and our measurement of $v\sin i$ to determine the line-of-sight spin axis inclination of the companion $i_p$.  However, simply computing the inclination as:

\begin{equation}
    \sin i = \bigg(\frac{v\sin i}{2\pi R/P_{\rm rot}}\bigg)
\end{equation}

\noindent does not account for correlations between relevant parameters \citep{Masuda2020}.  For example, $v$ and $v\sin i$ are not statistically independent given that $v\sin i$ is always less than $v$.  We therefore follow the method described in detail in \citet{Masuda2020} and summarized for the application to HD 106906 b below.

Given $v$ as the equatorial rotational velocity and $u$ = $v\sin i$ as the projected rotation rate, we have the following two likelihood functions:

\begin{align} 
L_v(v) = p(d_v | v) \\
L_u(u) = p(d_u | u) 
\end{align}

\noindent where $d_v$ and $d_u$ are the datasets from which these likelihood functions are calculated.  In our case, $L_u(u)$ is the probability distribution for $v\sin i$ that we determined from our high-resolution spectra, a Gaussian with peak location 9.5 km/s and standard deviation 0.2 km/s.  $L_v(v)$ is the probability distribution for $v$, which we calculate using $v$ = 2$\pi R$/$P_{\rm rot}$.  For the radius $R$ we calculate the effective blackbody radius:

\begin{equation}
    R = \sqrt{\frac{L}{4\pi \sigma_b T_{\rm eff}^4}},
\end{equation}

\noindent where $L$ is the bolometric luminosity log$(L_{\rm bol}/L_{\odot})$ = -3.65$\pm$0.08, $\sigma_b$ is the Stefan-Boltzmann constant, and $T_{\rm eff}$ is the effective temperature T$_{\rm eff}$ = 1820$\pm$240 K \citep{Daemgen2017}. This yields a radius of 1.49$^{+0.37}_{-0.45}$ R$_{\rm Jup}$. We produce a probability distribution for $v$ by incorporating uncertainties on $P_{\rm rot}$, log$(L_{\rm bol}/L_{\odot})$, and $T_{\rm eff}$ in a Monte Carlo fashion.

With $L_u(u)$ and $L_v(v)$ in hand, \citet{Masuda2020} specify two key assumptions:

1. $d_v$ and $d_u$ are independent, so the likelihood function for D = $\{d_v,d_u\}$ is separable:

\begin{equation}
\begin{split}
    L_{vu}(v,u) = p(D | v,u) = p(d_v | v, u) = \\
    p(d_v | v)p(d_u | u) = L_v(v)L_u(u)
\end{split}
\end{equation}

2.  $v$ and $i$ are a priori independent, which means that the prior P$_{vi}(v,i)$ is separable:

\begin{equation}
    P_{vi}(v,i) = P_v(v)P_i(i)
\end{equation}

\noindent and

\begin{equation}
    p(v | i) = P_v(v) ; p(i | v) = P_i(i)
\end{equation}

Given these assumptions, the posterior PDF for $\cos i$ can be written as:

\begin{equation}
    p(\cos i | D) \propto P_{\rm cosi}(\cos i)\int L_v(v)L_u(v\sqrt{1 - \cos^2 i})P_v(v) dv
\end{equation}

\noindent where $P_{\rm cosi}(\cos i)$ is uniform between 0 and 1 and the prior on the rotation rate $P_v(v)$ is uniform from 0 to break-up speed. 

\begin{figure}[h]
\centering
\includegraphics[width=0.5\textwidth]{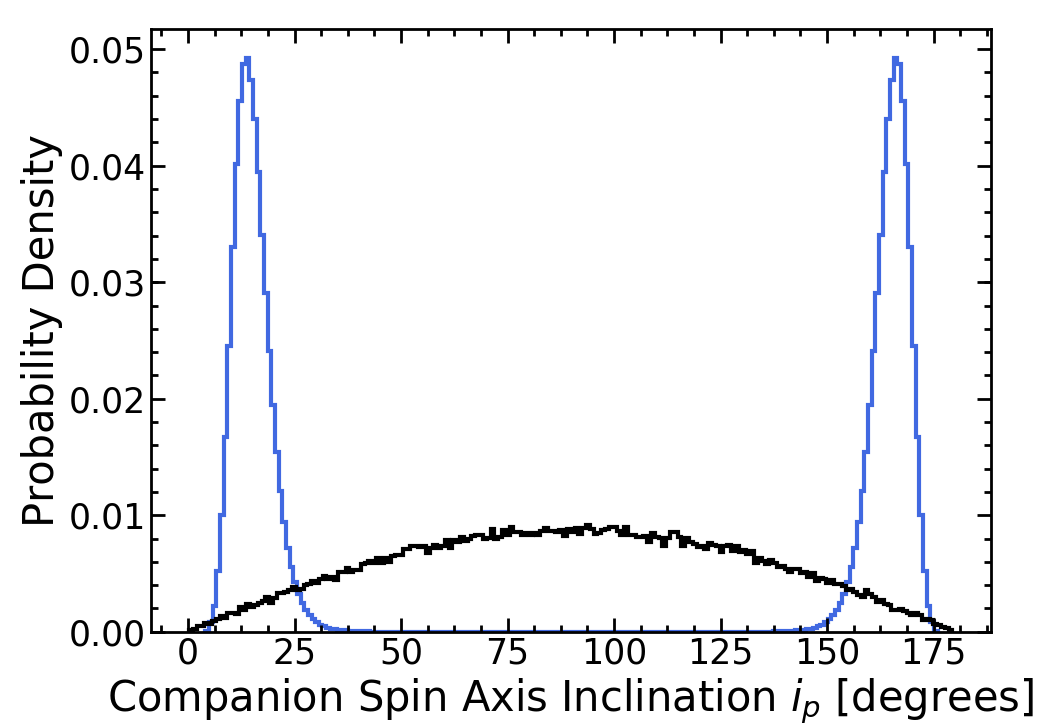}
\caption{Normalized posterior distribution of the line-of-sight companion spin axis inclination $i_p$ (blue). This distribution is bimodal and symmetric around 90$\degree$ because we do not know whether this spin axis vector (which has directionality) is pointed towards us or away from us.  Setting aside the symmetric distribution above 90$\degree$, just considering values of $i_p$ $<$ 90$\degree$ we find that the mode and 68$\%$ highest probability density interval of $i_p$ are 14 $\pm$ 4$\degree$. This distribution is compared to a random inclination distribution (black) whose values are drawn from a uniform distribution in $\cos i$.}
\label{fig: planetary spin axis posterior}
\end{figure}

Converting this PDF in $\cos i$ to a PDF in $i$ yields the distribution shown in Figure \ref{fig: planetary spin axis posterior}. We note that the posterior distribution for $i_p$ is bimodal and symmetric around 90$\degree$ simply because we do not know whether this spin axis vector (which has directionality) is pointed towards us or away from us.  Therefore the mode and 68$\%$ highest probability density interval (HPDI) of $i_p$ is 14 $\pm$ 4$\degree$ for $i_p$ $<$ 90$\degree$, and 166 $\pm$ 4$\degree$ for $i_p$ $>$ 90$\degree$.

\subsection{Measuring the 3D Spin-Orbit Architecture of the HD 106906 b System} \label{3D}

Our goal is to measure the true 3D angles between the three angular momentum vectors in this system -- the companion spin axis, the companion orbit normal, and the disk normal.  These angles are given by:

\begin{equation} \label{true_obl_op}
    \Psi_{op} = \cos^{-1}(\cos i_p\cos i_o + \sin i_p \sin i_o \cos (\Omega_o - \Omega_p))
\end{equation}
\begin{equation}\label{true_obl_dp}
    \Psi_{dp} = \cos^{-1}(\cos i_p\cos i_d + \sin i_p \sin i_d \cos (\Omega_d - \Omega_p))
\end{equation}
\begin{equation}\label{true_obl_od}
    \Psi_{od} = \cos^{-1}(\cos i_d\cos i_o + \sin i_d \sin i_o \cos(\Omega_o - \Omega_d))
\end{equation}

\noindent where $\Psi_{op}$ is the true companion obliquity, $\Psi_{dp}$ is the true spin-disk mutual inclination, and $\Psi_{od}$ is the true orbit-disk mutual inclination. The position angles $\Omega_o$, $\Omega_d$, and $\Omega_p$ measure how the orbit, disk, and companion spin axis, respectively, are oriented on the sky plane.  The nodal angle $\Omega_p$ is unknown.  

The difference between the line-of-sight inclination of the companion spin axis $i_p$ and that of the orbit normal $i_o$ yields a lower limit on the true de-projected obliquity $\Psi_{op}$ \citep{Bowler2017}:

\begin{equation} \label{obl_limit}
\Psi_{op} > |i_p - i_o| \,.
\end{equation}

\begin{figure}[h]
\centering
\includegraphics[width=0.45\textwidth]{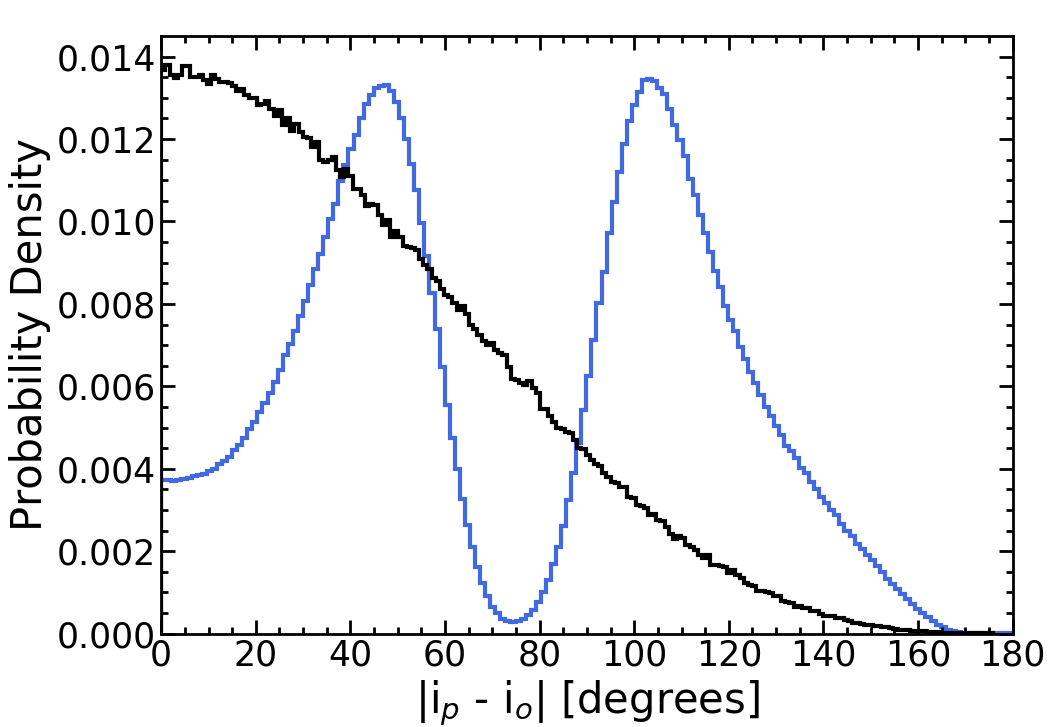}
\includegraphics[width=0.45\textwidth]{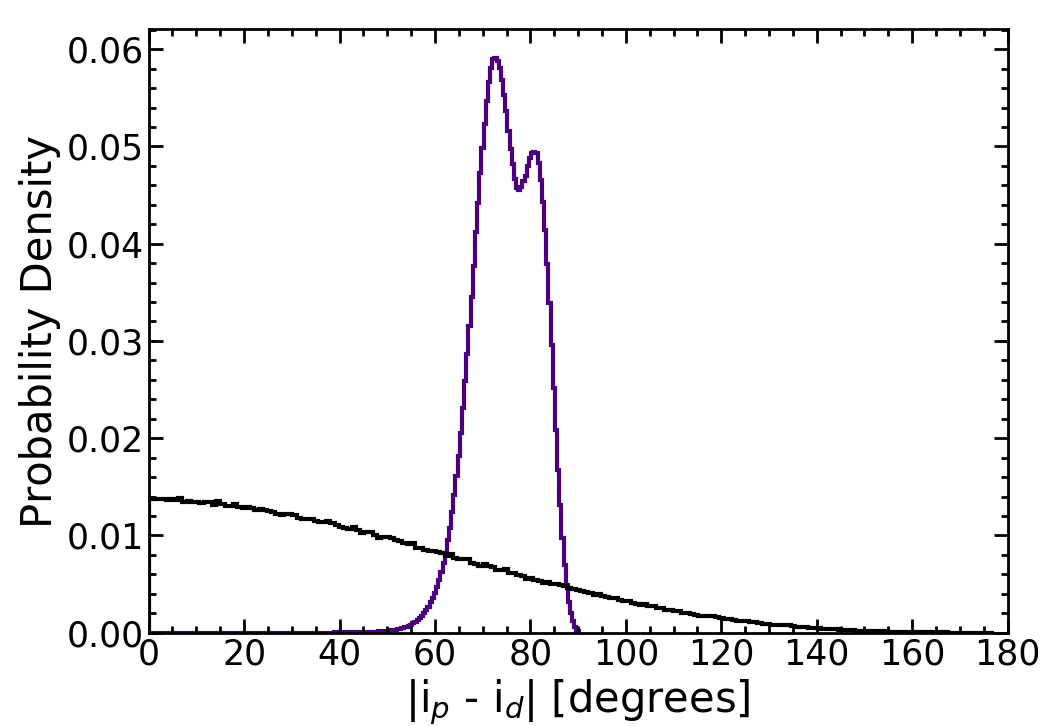}
\includegraphics[width=0.45\textwidth]{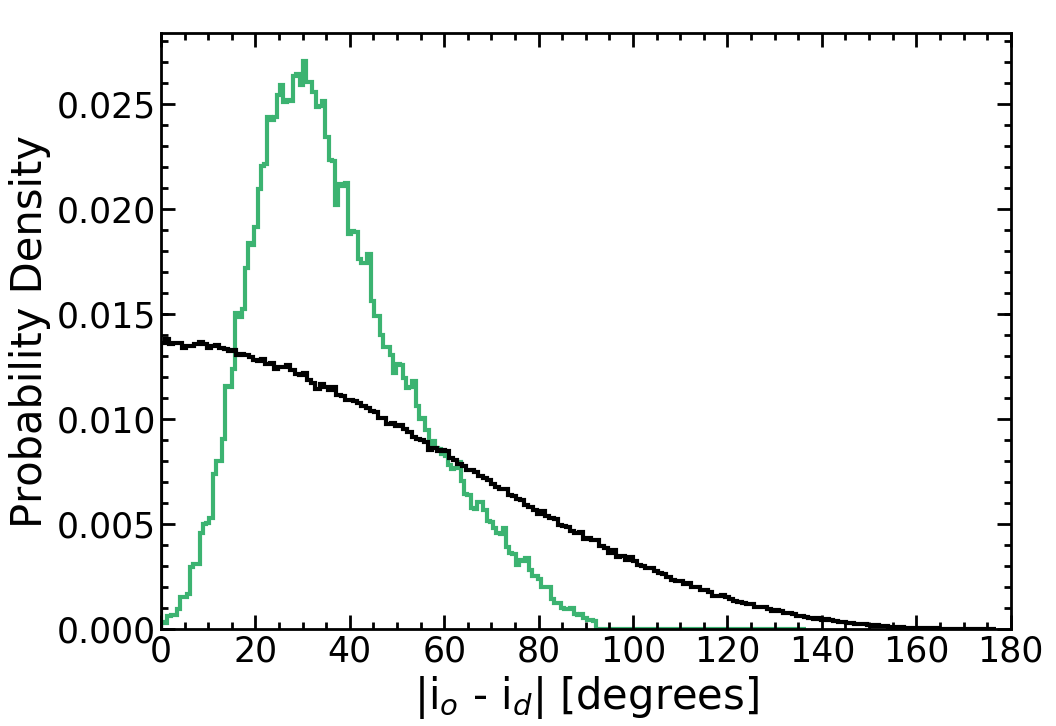}
\caption{Top panel: Posterior distribution for the line-of-sight projected companion obliquity (blue), which has a 68$\%$ highest probability density interval (HPDI) of [32, 119] degrees.  Middle panel:  Posterior distribution for the line-of-sight projected spin-disk angle (purple), which has a 68$\%$ HPDI of [69, 83] degrees.  Bottom panel:  Posterior distribution for the line-of-sight projected orbit-disk angle, which has a 68$\%$ HPDI of [16, 48] degrees.  These line-of-sight projections are lower limits on the true 3D mutual inclinations $\Psi_{op}$, $\Psi_{dp}$, and $\Psi_{od}$.  The posterior distributions are compared to a random projected inclination distribution (black), where $i_o$, $i_d$, and $i_p$ have all been drawn from uniform distributions in $\cos i$.}
\label{fig: projected obliquity posteriors}
\end{figure}

\noindent Similarly:

\begin{align}
\Psi_{dp} &> |i_p - i_d| \\
\Psi_{od} &> |i_o - i_d| \,.
\end{align}
Figure \ref{fig: projected obliquity posteriors} shows the posteriors for $|i_p - i_o|$, $|i_p - i_d|$, and $|i_o - i_d|$.  We also plot a random distribution in black for comparison, where $i_p$, $i_o$, and $i_d$ are all drawn from uniform distributions in $\cos i$.  We find that the 68$\%$ highest probability density interval (HPDI) for $|i_p - i_o|$ lies between [32, 119] degrees. For $|i_o - i_d|$ the 68$\%$ HPDI is [16, 48] degrees, and for $|i_p - i_d|$ we have the tightest 68$\%$ HPDI of [69, 83] degrees. 

\begin{figure}[h]
\centering
\includegraphics[width=0.5\textwidth]{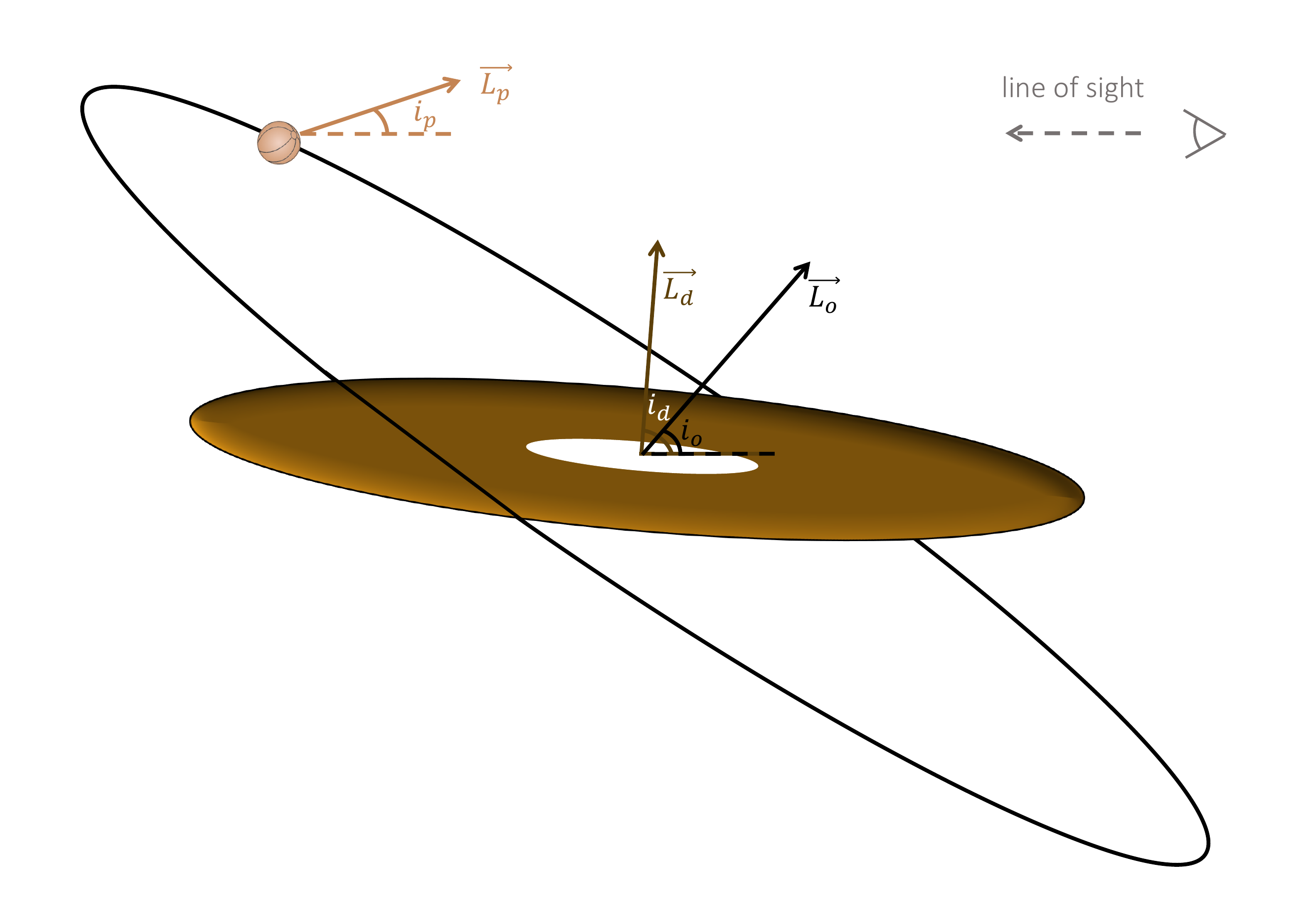}
\caption{3D architecture of the HD 106906 system showing how each of the three angular momentum vectors, namely $\vec{L}_p$ for the companion spin, $\vec{L}_o$ for the orbit, and $\vec{L}_{d}$ for the debris disk, are oriented along our line-of-sight, with corresponding angles $i_p$, $i_o$, and $i_d$. Both $i_p$ and $i_d$ are symmetric around 90$\degree$ ($\vec{L}_p$ and $\vec{L}_{d}$ could be pointing towards or away from us). Mutual inclinations between each set of vectors favor misalignment. Note the purpose of this figure is to illustrate the orientation of the three angular momentum vectors -- it does not accurately depict other separate system properties (i.e. disk asymmetry, disk inner and outer radius -- see Fig. 9 in \citet{Nguyen2021} for a scattered light image of the system showing some of these properties).}
\label{fig: 3D orientation}
\end{figure}

Since these projected angles are all lower limits on the true 3D angles, we can say that our results tend to favor more ``misaligned'' orientations (defined here as mutual inclinations of 20--180$\degree$). A schematic illustration of the line-of-sight architecture of the system is shown in Figure \ref{fig: 3D orientation}.

\begin{deluxetable*}{ccc}
\tablecaption{ Measured Parameters \label{tb:obl}}
\tabletypesize{\footnotesize}
\tablehead{
  \colhead{Parameter} & 
  \colhead{Measured Value} & 
  \colhead{Ref}
}
\startdata
$v_p \sin i_{p}$ & 9.5$\pm$0.2 km/s & This work \\
$P_{rot,p}$ & 4.1$\pm$0.3 hrs & \citet{Zhou2020} \\
$i_p$ & 14$\pm$4 or 166$\pm$4 deg & This work \\
$i_o$ & 56$^{+12}_{-21}$ deg & \citet{Nguyen2021} \\
$i_d$ & 85.3$\pm$0.1 or 94.7$\pm$0.1 deg & \citet{Kalas2015,Lagrange2016,Nguyen2021} \\
$|i_o - i_p|$ & [32, 119] deg & This work \\
$|i_p - i_d|$ & [69, 83] deg & This work \\
$|i_d - i_o|$ & [16, 48] deg & This work \\
$\Psi_{op}$ & 55$^{+22}_{-16}$ or 125$^{+16}_{-22}$ deg & This work \\
$\Psi_{dp}$ & 84$^{+6}_{-8}$ or 96$^{+8}_{-6}$ deg & This work \\
$\Psi_{od}$ & 39$^{+20}_{-15}$ or 141$^{+15}_{-20}$ deg & This work
\enddata
\tablecomments{The three $i$ inclinations presented here are all along our line-of-sight.  The angle $i_p$ is symmetric about 90$\degree$ due to the fact that we do not know whether this spin angular momentum vector is pointing towards us or away from us. The angle $i_d$ has two solutions because it is unclear whether the disk is rotating prograde or retrograde, so there are two combinations of $i_d$ and $\Omega_d$, where $\Omega_d$ is the position angle of the ascending node, that are equally likely.  The line-of-sight mutual inclinations $|i_o - i_p|$, $|i_p - i_d|$, and $|i_d - i_o|$ are lower limits on the true de-projected angles $\Psi_{op}$, $\Psi_{dp}$, and $\Psi_{od}$.  Here we quote the mode and 68$\%$ highest probability density intervals for these angles.}
\end{deluxetable*}

\begin{figure}[h]
\centering
\includegraphics[width=0.5\textwidth]{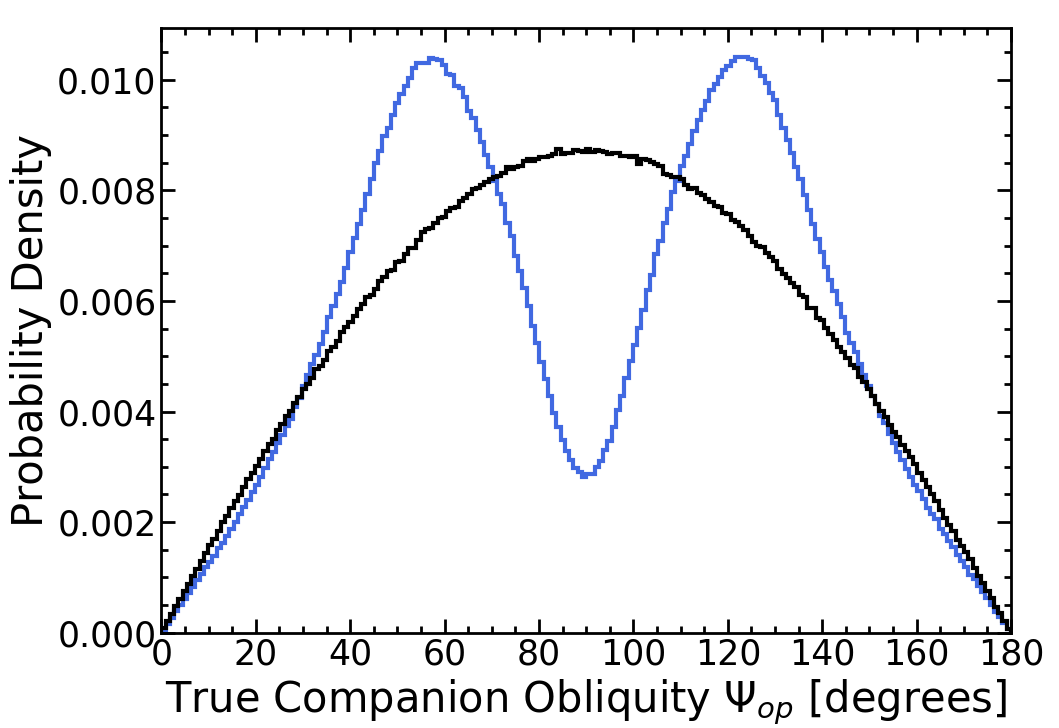}
\includegraphics[width=0.5\textwidth]{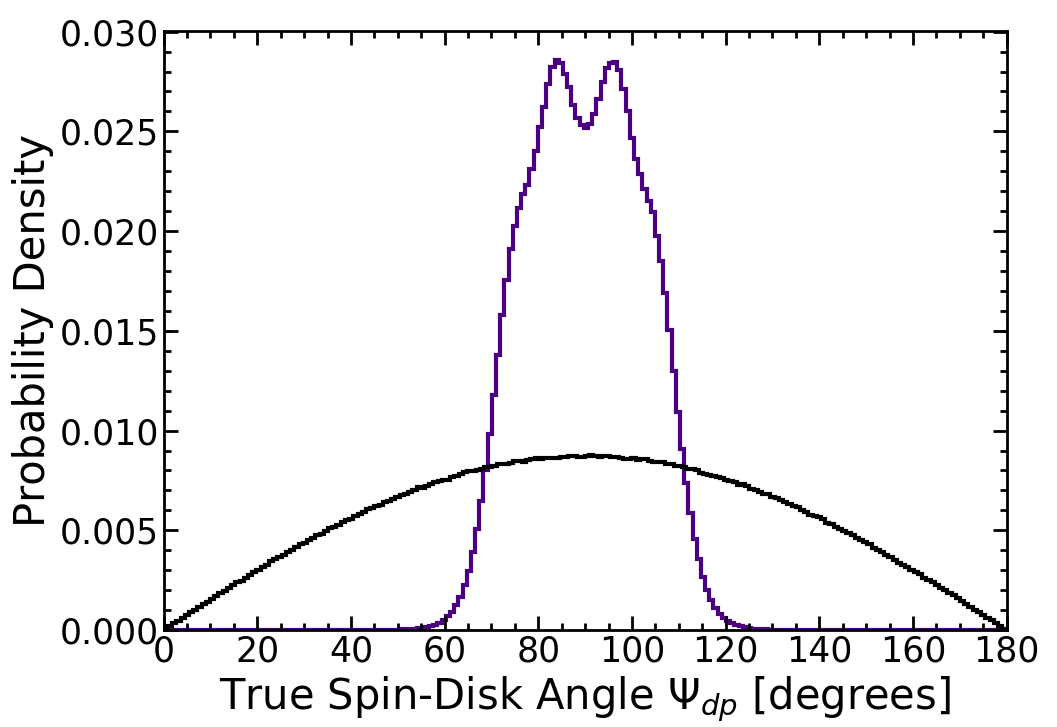}
\includegraphics[width=0.5\textwidth]{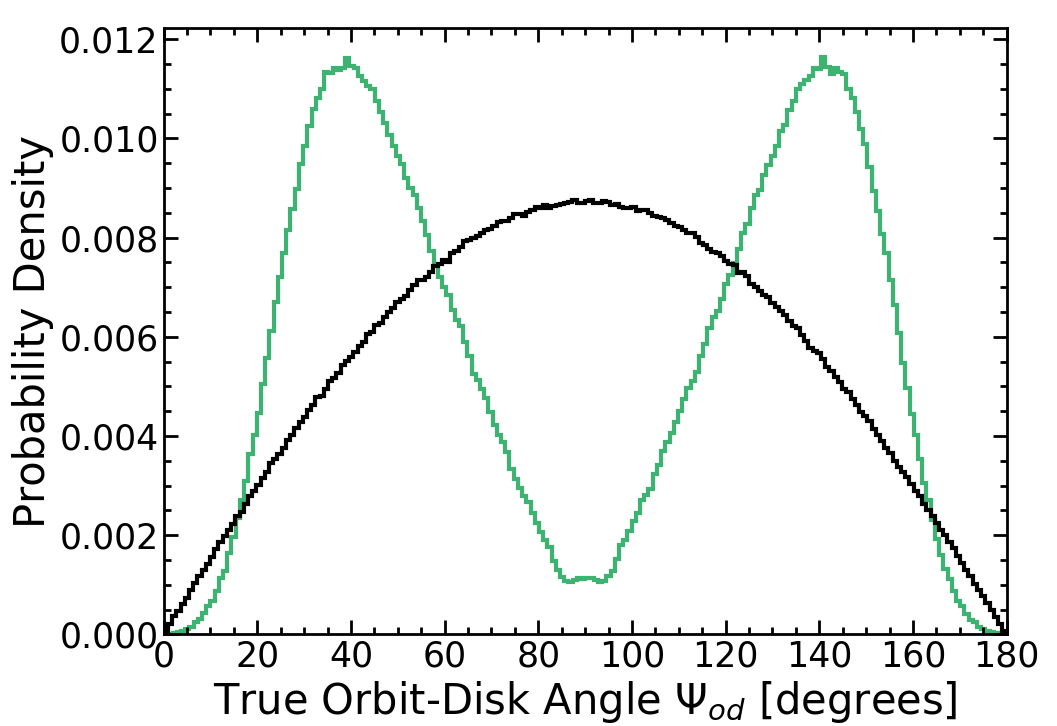}
\caption{Top panel: Normalized posterior distribution for the true companion obliquity $\Psi_{op}$ (blue).  Middle panel:  Normalized posterior distribution for the true spin-disk angle $\Psi_{dp}$ (purple).  Bottom panel:  Normalized posterior distribution for the true orbit-disk angle $\Psi_{od}$ (green).  These posterior distributions are compared to geometrically random mutual inclination distributions (black).}
\label{fig: true obliquity posteriors}
\end{figure}

We now calculate full probability distributions for all $\Psi$'s using equations  (\ref{true_obl_op})-(\ref{true_obl_od}).  For equations (\ref{true_obl_op}) and (\ref{true_obl_dp}) we assume $\Omega_p$ is randomly drawn from a uniform distribution between 0 and 2$\pi$.  For equation (\ref{true_obl_od}), both $\Omega_o$ and $\Omega_d$ have been measured:  $\Omega_o$ = 99$^{+26}_{-28}$$\degree$ or 279$^{+25}_{-29}$$\degree$, and $\Omega_d$ = 104.4 $\pm$ 0.3$\degree$ or 284 $\pm$ 0.3$\degree$ \citep{Kalas2015,Lagrange2016,Nguyen2021}.  Figure \ref{fig: true obliquity posteriors} shows the resulting probability distributions for $\Psi_{op}$, $\Psi_{dp}$, and $\Psi_{od}$, along with a random mutual inclination distribution $\Psi_{\rm random}$ in black for reference. 

Not knowing $\Omega_p$ along with poor constraints on HD 106906 b's orbital elements (a consequence of the companion's wide orbital separation) leads to broad posterior distributions for the true de-projected angles $\Psi_{op}$ and $\Psi_{od}$. By comparison, $\Psi_{pd}$ is remarkably well-constrained. All of these posteriors are bimodal, reflecting symmetries across 90$\degree$. 
For each $\Psi$ posterior we calculate the mode and 68$\%$ HPDI for each half of the distribution below and above 90$\degree$.  We find that the true companion obliquity $\Psi_{op}$ is 55$^{+22}_{-16}$$\degree$ or 125$^{+16}_{-22}$$\degree$, the true spin-disk angle $\Psi_{dp}$ is 84$^{+6}_{-8}$$\degree$ or 96$^{+8}_{-6}$$\degree$, and the true mutual inclination between the orbit and disk normals $\Psi_{od}$ is 39$^{+20}_{-15}$$\degree$ or 141$^{+15}_{-20}$$\degree$ (Table \ref{tb:obl}).  Each of these de-projected mutual inclinations deviates distinctly from the geometric prior -- both $\Psi_{op}$ and $\Psi_{od}$ favor angles away from 90$\degree$ which is where the geometric prior peaks, and while $\Psi_{dp}$ peaks around 90$\degree$ it is a much tighter constraint than a random distribution.

We quantify the probability that each $\Psi$ distribution yields an ``aligned'' state, which we define as $\Psi \in (0,20)$ degrees, or a ``misaligned'' state where $\Psi \in (20,180)$ degrees. The Bayesian odds ratio is $p(m | D)/p(a | D)$, where $p(m | D)$ is the probability of misaligned state $m$ given data $D$, and $p(a | D)$ is the probability of an aligned state $a$. The former probability is the integral of the posterior distribution $p(\Psi | D)$ (the colored histograms in Fig. \ref{fig: true obliquity posteriors}) from 20 - 180$\degree$, while the latter probability is obtained by integrating the posterior distribution from 0 - 20$\degree$.

We compute these integrals over the posterior distributions for all three mutual inclinations $\Psi_{op}$, $\Psi_{dp}$, and $\Psi_{od}$, as well as for the geometric prior distribution which is uniform in $\cos \Psi_{\rm random}$ (Fig.~\ref{fig: true obliquity posteriors}, black curve).  In the absence of any data, the geometric prior yields an odds ratio favoring misalignment to alignment at 32:1 (2.2$\sigma$).  The odds ratios for the true companion obliquity $\Psi_{op}$ and the orbit-disk mutual inclination $\Psi_{od}$ are only slightly larger, 39:1 (2.3$\sigma$) and 40:1 (2.3$\sigma$), respectively. Nevertheless, the fact that the shapes of both posteriors differ significantly from that of the prior indicates that the odds favoring misalignment for both $\Psi_{op}$ and $\Psi_{od}$ are not entirely driven by the prior. We can say there is tentative evidence that both of these mutual inclinations are large. By comparison, the true spin-disk angle $\Psi_{dp}$ has an odds ratio that favors misalignment much more strongly than does our random prior -- 97081:1 (4.4$\sigma$).  We can say with confidence that this system contains an edge-on disk and a planet spinning nearly pole-on; even allowing for sky-plane degeneracy, our measurements indicate that the companion spin axis and disk normal are strongly misaligned.

\section{Discussion: Possible Formation Histories}

For the HD 106906 system, we have strong evidence that the spin axis of the planetary-mass companion (PMC) and the orbit normal of the debris disk are misaligned. In addition, we find tentative evidence that the true PMC obliquity, and the mutual inclination between the PMC orbit and debris disk, are large. We proceed on the assumption that all three vectors are mutually misaligned and consider possible origin scenarios.

As in the case of 2M0122 b, gravitational instability in a turbulent environment is a promising way to create tilted architectures. The setting can either be a gravito-turbulent disk around a protostar \citep{Bryan2020,Jennings2021}, or a portion of a self-gravitating turbulent cloud that fragments into a binary \citep[e.g.][]{bate2002,bate2009,Bate2018}. Turbulent eddies spin in a variety of directions, and when those eddies gravitationally collapse, the objects they form should have a correspondingly wide range of spin directions. In situ formation of HD 106906 b in a disk would require a nebula as large as 1000 au in radius; Class I disks range up to this size (\citealt{Maury2019}, see their Figure 9). Numerical simulations of fragmenting gravito-turbulent disks by \citet{Jennings2021} find that obliquities of nascent fragments can be as high as 45$\degree$, and that subsequent collisions between fragments can raise or lower obliquities. Turbulence also imparts vertical velocities to disk fragments, driving them out of the midplane; orbital inclinations up to 20$^\circ$ are suggested by these simulations. 
In population synthesis calculations by \citet{Bate2018} of stellar binaries fragmenting from a turbulent cloud, roughly 2/3 of circumstellar disks are inclined by more than 30$\degree$ relative to the binary orbit, for binaries with semimajor axes between 100 and 1000 au (see their Figure 20). This result of frequent disk-orbit misalignments, in combination with their finding that most (80\%) of stellar spin axes are within $\sim$45$\degree$ of circumstellar disk normals (their Figure 23),
would seem to imply that spin-orbit angles are typically large (obliquities are not explicitly calculated in \citealt{Bate2018}). In sum, all the misalignments indicated by our observations seem possible to account for by turbulent gravitational instability. 

Another scenario is that the PMC formed as an isolated object via molecular cloud fragmentation and was subsequently captured by a star. Binaries can form in fly-by events if sufficient energy is dissipated during the encounter; circumstellar disks can provide this energy sink \citep[e.g.][]{Clarke1991,Clarke19912,Moeckel2006,Offner2016,Bate2018}. Assuming the disk surrounding the stellar host is the dominant sink (as opposed to any smaller and less massive disk orbiting the PMC), we have no reason to expect the planet's spin axis to be aligned with its orbit normal or the disk normal. The captured PMC's orbital plane may also be randomly oriented with respect to the circumstellar disk plane, at least initially; this is borne out in simulations by \citet[][see their Figure 22]{Bate2018}. However, these simulations lasted only up to $\sim$$10^5$ yr; on longer timescales the PMC orbit could be gravitationally torqued into alignment with the circumstellar disk. Thus two if not three misalignments can be accommodated in this encounter scenario.

A handful of other mechanisms can produce only one or two of the three possibly large mutual inclinations observed in this system. A stellar fly-by can tilt the orbit of the PMC, generating a large obliquity as well as a mutual inclination between the orbit and disk planes \citep[e.g.][]{Laughlin1998,Kenyon2004,Zakamska2004,Parker2012,deRosa2019}.  However, the PMC spin axis is not expected to be materially altered by the fly-by, and thus the observed misalignment between the spin axis and the disk normal in HD 106906 is unexplained in this scenario. Another way to generate obliquity is by using the stellar tidal potential to tilt a circumplanetary disk and by extension its host planet \citep{Lubow2000,Martin2021}. This scenario does not address the misalignment between the orbit and disk normals.

\section{Conclusions}

In this study we constrained the orientation of the planetary spin, orbital, and disk angular momentum vectors for the HD 106906 system.  HD 106906 is a 13 $\pm$ 2 Myr system in ScoCen composed of a close binary of two F stars, a widely separated planetary-mass tertiary 106906 b, and an asymmetric debris disk interior to companion b's orbit. Line-of-sight inclinations for the companion orbit $i_o$ and debris disk $i_d$ were previously published \citep{Kalas2015,Lagrange2016,Nguyen2021}. The debris disk is known to be viewed edge-on. Here we measured the line-of-sight inclination of the companion spin axis $i_p$.  We used near-IR high-resolution spectra from IGRINS/Gemini South to measure rotational line broadening for HD 106906 b, and found a rotation rate of 9.5 $\pm$ 0.2 km/s. Combining this measurement with the photometric rotation period yielded a companion spin axis inclination of 14 $\pm$ 4$\degree$ (when considering inclinations $<$ 90$\degree$). We are seeing this companion nearly pole-on (Fig. \ref{fig: 3D orientation}).  

Differences between line-of-sight inclinations yield lower limits on the true 3D mutual inclinations.  We computed the projected inclinations $|i_o - i_p|$, $|i_p - i_d|$, and $|i_d - i_o|$. The projected companion obliquity $|i_o - i_p|$ has a 68$\%$ highest density probability interval (HDPI) of [32, 119] degrees. The projected orbit-disk angle $|i_d - i_o|$ has an HDPI of [16, 48] degrees. 
The projected spin-disk inclination $|i_p-i_d|$ is the most strongly constrained, with an HDPI of [69, 83] degrees.  These lower limits on the 3D mutual inclinations favor more ``misaligned'' orientations (defined here as mutual inclinations of 20--180 degrees).

We further constrained the true 3D mutual inclinations between the companion spin, companion orbit, and disk angular momentum vectors: $\Psi_{op}$ (true companion obliquity), $\Psi_{dp}$ (true spin-disk mutual inclination), and $\Psi_{od}$ (true orbit-disk angle).  Since we have no constraints on the companion spin axis orientation in the sky plane, to compute distributions for $\Psi_{op}$ and $\Psi_{dp}$ we assumed that nodal angle $\Omega_p$ was randomly and uniformly distributed between 0 and $2\pi$ (see equations \ref{true_obl_op} and \ref{true_obl_dp}).  To calculate a distribution for $\Psi_{od}$ we used measured values of $\Omega_d$ and $\Omega_o$ for the disk and orbit, respectively (see equation \ref{true_obl_od}). 

The lack of a constraint on how the companion spin axis is oriented in the sky plane combined with poor constraints on HD 106906 b's orbit (given its wide projected separation) leads to broad posterior distributions for $\Psi_{op}$ and $\Psi_{od}$, whereas $\Psi_{dp}$ is more tightly constrained (Figure \ref{fig: true obliquity posteriors}). Since these posteriors are bimodal (given symmetries about 90$\degree$), we calculated the mode and 68$\%$ HPDI for each $\Psi$ both above and below 90$\degree$. We found that $\Psi_{op}$ is 55$^{+22}_{-16}$$\degree$ or 125$^{+16}_{-22}$$\degree$, $\Psi_{dp}$ is 84$^{+6}_{-8}$$\degree$ or 96$^{+8}_{-6}$$\degree$, and $\Psi_{od}$ is 39$^{+20}_{-15}$$\degree$ or 141$^{+15}_{-20}$$\degree$. All three angles exhibit marked differences from the geometric prior, with $\Psi_{op}$ and $\Psi_{od}$ peaking at values away from the 90 degree maximum of the geometric prior, and $\Psi_{dp}$ showing a much tighter distribution around 90 degrees than the prior.

We assessed how likely the three angular  momentum vectors were to be ``misaligned'' or ``aligned'' (defined as mutual inclinations between 20--180$\degree$ and 0--20$\degree$ respectively) by calculating an odds ratio for each angle.  Both the true obliquity $\Psi_{op}$ and mutual inclination between the orbit and the disk $\Psi_{od}$ tentatively favor misalignment with ratios of 39:1 (2.3$\sigma$) and 40:1 (2.3$\sigma$), respectively.  The orientation between the disk normal and the companion spin axis $\Psi_{dp}$ strongly favors misalignment at 97081:1 odds  (4.4$\sigma$). We are seeing an edge-on disk orbited by a planet spinning nearly pole-on.

Given strong evidence that the planetary-mass companion (PMC) spin axis and disk normal are misaligned, and tentative evidence that the true PMC obliquity and the mutual inclination between the PMC orbit and debris disk are large, we considered various origin scenarios.  As in the case of 2M0122 b \citep{Bryan2020}, gravitational instability in a turbulent medium, either in a circumstellar disk or cloud setting, is viable.  Gravitational collapse of turbulent eddies naturally yields large obliquities, and subsequent fragment interactions (collisions and mergers) can further increase obliquity dispersions \citep{Jennings2021}. Disk turbulence also excites non-zero orbital inclinations. In simulations of binary star fragmentation in a turbulent molecular cloud, disks and binary orbits are frequently misaligned, especially for wide binaries \citep{Bate2018}. Thus all observed misalignments in HD 106906 are potentially accounted for in a turbulent gravitational instability scenario.  Another possibility is that the PMC formed via molecular cloud fragmentation initially isolated and unbound from the star, and was subsequently captured by the star in a dissipative fly-by event. In this encounter scenario there is no reason to expect the PMC's spin axis would be aligned with its orbit normal or the disk normal, and the orbital plane could also be randomly oriented with respect to the disk plane.  
In all of the above scenarios, HD 106906 b forms top-down by gravitational instability, and thus shares kinship with stars.

This is only the second obliquity constraint for a planetary-mass companion outside the Solar System.   
The work of measuring a planet's rotation speed, spin period, and 3D orbit remains challenging. But the insights into planet formation enabled by obliquity are new, powerful, and unique. 
The bane of sky-plane degeneracies may be banished
by a larger sample of systems like 2M0122 b and HD 106906 b which will enable statistical
constraints.

\section*{}
We thank Ian Czekala, Gaspard Duchene and Yifan Zhou for helpful conversations.  M.L.B. is supported by the Heising-Simons Foundation 51 Pegasi b Fellowship.  C.V.M. acknowledges the support of the National Science Foundation grant number 1910969. B.P.B. acknowledges support from the National Science Foundation grant AST-1909209 and NASA Exoplanet Research Program grant 20-XRP20$\_$2-0119.

Based on observations obtained at the international Gemini Observatory, a program of NSF’s NOIRLab, which is managed by the Association of Universities for Research in Astronomy (AURA) under a cooperative agreement with the National Science Foundation. on behalf of the Gemini Observatory partnership: the National Science Foundation (United States), National Research Council (Canada), Agencia Nacional de Investigaci\'{o}n y Desarrollo (Chile), Ministerio de Ciencia, Tecnolog\'{i}a e Innovaci\'{o}n (Argentina), Minist\'{e}rio da Ci\^{e}ncia, Tecnologia, Inova\c{c}\~{o}es e Comunica\c{c}\~{o}es (Brazil), and Korea Astronomy and Space Science Institute (Republic of Korea).
This work used the Immersion Grating Infrared Spectrometer (IGRINS) that was developed under a collaboration between the University of Texas at Austin and the Korea Astronomy and Space Science Institute (KASI) with the financial support of the Mt. Cuba Astronomical Foundation, of the US National Science Foundation under grants AST-1229522 and AST-1702267, of the McDonald Observatory of the University of Texas at Austin, of the Korean GMT Project of KASI, and Gemini Observatory.


\bibliographystyle{aasjournal}
\bibliography{bibliography}



\end{document}